\documentclass[%
 reprint,
superscriptaddress,
showpacs,
 amsmath,amssymb,
 aps,
 prl,
 floatfix,
]{revtex4-2}

\usepackage{graphicx}
\usepackage{dcolumn}
\usepackage{bm}
\usepackage{hyperref}
\usepackage[mathlines]{lineno}
\usepackage{xcolor}
\pagecolor{white}

\begin{document}
\title{Intense multicycle THz pulses generated by laser-produced nanoplasmas}


\author{Manoj Kumar}
\affiliation{Department of Physics, Ulsan National Institute of Science and Technology, Ulsan 44919, Republic of Korea}
\author{Hyung Seon Song}
\affiliation{Department of Physics, Ulsan National Institute of Science and Technology, Ulsan 44919, Republic of Korea}
\author{Jaeho Lee}
\affiliation{Department of Physics, Ulsan National Institute of Science and Technology, Ulsan 44919, Republic of Korea}
\author{Dohyun Park}
\affiliation{Department of Physics, Ulsan National Institute of Science and Technology, Ulsan 44919, Republic of Korea}
\author{Hyyung Suk}
\affiliation{Department of Physics and Photon Science, Gwangju Institute of Science and Technology, Gwangju 61005, Republic of Korea}
\author{Min Sup Hur}
\email[]{mshur@unist.ac.kr}
\affiliation{Department of Physics, Ulsan National Institute of Science and Technology, Ulsan 44919, Republic of Korea}


\date{\today}

\begin{abstract}
We present a novel scheme to obtain robust, narrowband, and tunable THz emission by using a nano-dimensional overdense plasma target that is irradiated by two counter-propagating detuned laser pulses. So far, no narrowband THz sources with a field strength of GV/m-level have been reported from laser-solid interaction. We report intense THz pulses at beat-frequency ($\simeq$30THz) generated due to strong plasma current produced by the beat ponderomotive force in colliding region, with an unprecedentedly high peak field strength of 11.9GV/m and spectral width $(\Delta f/f$$\simeq$$5.3\%)$ from 2D PIC simulations. Such an extremely bright narrowband THz source is suitable for various ambitious applications.
\end{abstract}


\maketitle


In recent years, terahertz (THz) radiation sources have identified potential in a variety of research and industrial fields including biomedical imaging, time-resolved molecular spectroscopy, communication, and security \cite{Tonouchi1, Zhang2, Pickwell3, Mittleman4, Jepsen5, Baxter6, Liu7, Koenig8, Kemp9}. Despite the rapid advances in THz science today, developing intense and compact THz radiation sources still remains challenging. Such intense THz sources would open up a wide range of research areas. 

Great strides have been made in crystal-based THz devices \cite{Auston10, Stepanov11, Fulop12, Huang13, Vicario14, Vicario15} that generate sub-millijoule energies of THz pulses with conversion efficiencies reaching around $1\%$. However, it is very difficult to scale up to higher energies since the crystals can be damaged at high intensities of driving laser pulses. By contrast, plasma can overcome the problem of optical damage to crystals, because it can sustain extremely high-amplitude electromagnetic oscillations. Such characteristics of plasmas have attracted considerable attention as intense and powerful THz sources \cite{Hamster16, Sheng17, Kumar18, Kwon19, Kumar20}.  Mechanism-wise, ionization current in laser-induced plasma filaments can generate THz pulses with wide spectral bandwidth and microjoules pulse energy \cite{Amico21, Kim22, Dai23, Liu24, Kuk25}. Another popular mechanism is using coherent transition radiation (CTR) from a solid target, where intense THz pulses with 1GV/m-level, half-cycle pulses can be generated \cite{Leemans26, Schroeder27, Liao28, Liao29, Dechard30, Ding31}. Sheath radiation (SR) mechanism by transient sheath field associated with energetic ions was also reported \cite{Gopal32, Gopal33}. Many experiments of THz emission from laser-solid target interactions have been performed with tens of millijoules and conversion efficiency $0.1 \%$ \cite{Li34, Li35, Jin36, Herzer37, Liao38, Liao39}. Recently, structured targets were used to enhance the THz emission by increased laser absorption \cite{Mondal40, Tokita41, Zhang42}.

The systems described above generate very intense and high-power, but mostly half- or single-cycle THz pulses, which have a broad frequency spectrum. The broadband pulses are useful for ultrafast pump-probe experiments where THz pulse can be used as a strong DC electric bias field or spectroscopic study of materials \cite{Spies43}. However, spectral density at a particular, desired frequency is inevitably low, hence they are not suitable for applications that require monochromatic THz waves, \textit{e.g.} THz-driven accelerator \cite{Nanni44} or pump-probe experiments for molecular bonding \cite{Forst45, Dienst46, Bakke47}. For plasma-based narrowband THz sources, plasma oscillation \cite{Sheng17} \cite{Kwon19} can be used as a radiating antenna, but low coupling efficiency between plasma wave and THz radiation matters. Another method is generating THz emission at beating frequency of two laser pulses that co-propagate through underdense plasmas \cite{Kumar20}. In this case, conversion of beat-to-THz is technically difficult and inefficient, and most of all, plasma current that is responsible for THz emission is weak in underdense plasmas. Other than those ideas, free-electron laser is only a narrowband source in a high-power regime, but a large facility is required. 

In this work, we introduce a novel way to overcome the low plasma current and technical complexity in beat-wave schemes. In our idea,  two laser pulses counter-propagate (in contrast to conventional co-propagating schemes) and collide on a nano-dimensional overdense plasma sheet target. As the counter-propagating laser pulses interact with each other on a nano-thickness plasma, the restriction of \textit{low-density} can be eliminated since the pulses need not propagate inside the plasma. Large plasma current becomes available as the charge carrier density is high in overdense plasmas. No complicated conversion mechanism of beat-to-THz is required; the beat current can emit the THz wave directly into the vacuum through a thin (below the skin depth) nano-plasma sheet. No plasma shield of electromagnetic emission leads to high conversion efficiency. The number of oscillation cycles in THz pulses can be made arbitrarily large by increasing the duration of driving pulses. From these features, extremely intense multicycle THz pulses with a narrowband frequency spectrum at the beat-frequency of the laser pulses can be obtained with a high efficiency. Our two-dimensional particle-in-cell simulations show that such a THz source is capable of providing field strength of GV/m-level, and frequency tunability with a narrowband spectrum. Furthermore, our system preserves the advantage of a single-pulse-driven target, as the broadband THz radiation is emitted together with the desired narrowband one in a single setup. 

\begin{figure}
\includegraphics[scale=0.72]{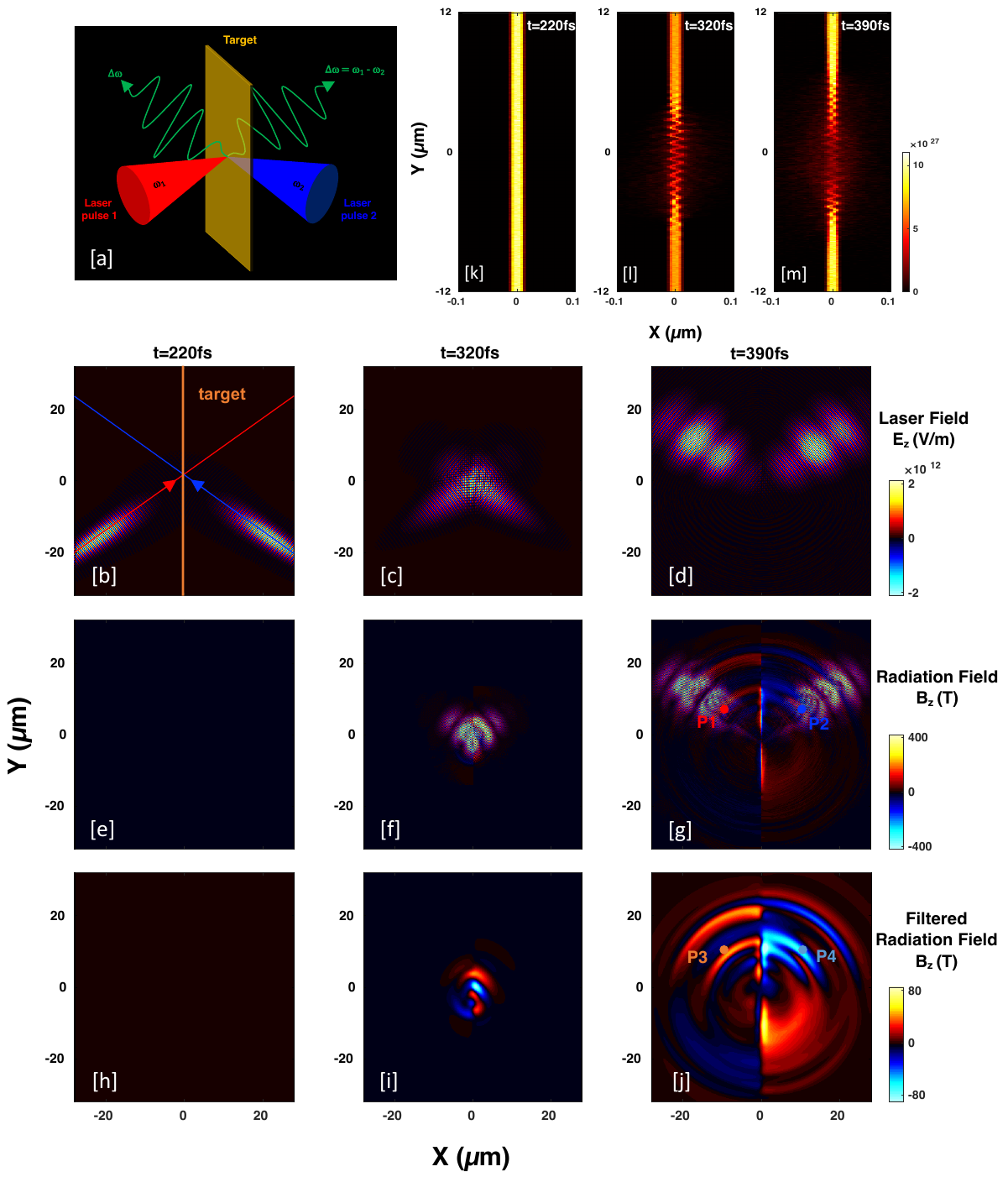}
\caption{\label{fig:fig1} 
(a) Sketch of the scheme: Two laser pulses with different frequencies, $\omega_{1}$ and $\omega_{2}$ collide obliquely with each other on an overdense plasma sheet of $nm$ thickness. From the region of the colliding laser pulses, beat-frequency radiation $(\Delta\omega$$=$$\omega_{1}$$-$$\omega_{2})$ is emitted into the vacuum. 
Snapshots of spatial distributions of; 
(b)-(d) laser field $E_{z} (\text{V}/\text{m})$ (s-polarized), 
(e)-(g) total radiation field $B_{z} (\text{T})$ (p-polarized),  
(h)-(j) filtered radiation field $B_{z} (\text{T}$), and 
(k)-(m) electron number density $n_{e} (\text{m}^\text{-3})$, at different times; $t$$=$$220\text{fs}$, $320\text{fs}$, and $390\text{fs}$.}
\end{figure}

For the simulation studies, we used the EPOCH code \cite{Arber48}.  The sketch of our THz scheme is shown in Fig.~\ref{fig:fig1}(a). Two laser pulses with different frequencies, $\omega_{1}$$=$$2\pi c/\lambda_{L1}$ and $\omega_{2}$$=$$2\pi c/\lambda_{L2}$, where the wavelengths $\lambda_{L1}$$=$$\text{800nm}$ and $\lambda_{L2}$$=$$\text{740nm}$, respectively, collide obliquely with each other at an angle $\theta_{\text{col}}$$=$$35^\circ$ on a plasma sheet target. The initial thickness and electron density of the plasma sheet are $d$$=$$\text{20nm}$ and $n_{e0}$$=$$5n_{c}$, respectively, where $n_{c}$$=$$m_{e}\varepsilon_{0}\omega_{1}^{2} /e^{2}$ is the critical density, $\varepsilon_{0}$ is the free space permittivity, $m_{e}$ and $-e$ are the electron mass and charge. The simulation domain is $56\mu \text{m}$ long and  $64\mu \text{m}$ wide with a grid size $dx$$=$$dy$$=$$\lambda_{L1}$/160 in the $x$- and $y$-directions, respectively. Each grid cell is filled with 40 numerical macro-particles. In our simulations, both laser pulses are injected into the simulation domain from the left and right boundaries, and they collide at the center of the plasma sheet. The laser pulses are s-polarized (to easily distinguish them from a p-polarized THz emission). The peak value of normalized vector potential of the lasers is $a_{1,2}$$=$$0.6$ (corresponding to intensities $I_{L1}$$\simeq$$7.7\times10^{17}\text{W}/\text{cm}^{2}$ and $I_{L2}$$\simeq$$9.5\times 10^{17}\text{W}/\text{cm}^{2}$), width size $b_{w}$$=$$5.8 \mu \text{m}$ and duration $\tau_{L}$$=$$100 \text{fs}$ (FWHM).

As both laser pulses interact with the plasma sheet, their beating exerts a ponderomotive force (PF) on the electrons of the plasma sheet that produces a nonlinear current, which emits the beat-frequency radiations $(\Delta\omega$$=$$\omega_{1}$$-$$\omega_{2})$ in THz frequency range. Along with the THz radiation, the second-harmonics (SH) of the incident lasers are also generated due to the second-harmonic current that is produced by the second-harmonic component of ponderomotive force. The snapshots of the electric field $E_{z}$ (it corresponds to the laser electric field for the s-polarized laser pulses), and the snapshots of the magnetic field $B_{z}$ (it corresponds to the radiation field; the model indicates that it is always p-polarized to the laser incident plane, regardless of the laser polarization) are presented in Figs.~\ref{fig:fig1}(b)-(d) and Figs.~\ref{fig:fig1}(e)-(g), respectively. The source of the emission is clearly at the colliding region of the laser pulses. The low-frequency part of this radiation could be isolated by applying a Gaussian-Filter to the total radiation field as shown in Figs.~\ref{fig:fig1}(h)-(j). As the beat ponderomotive force drives electron oscillations, the plasma electron density is strongly modulated in the colliding region at  $t=320\text{fs}$, as shown in Figs.~\ref{fig:fig1}(k)-(m). To determine the frequency of the emitted radiation in vacuum, we located virtual probes at P1($x$=$-10\mu\text{m}$, $y$=$7\mu\text{m}$), P2($x$=$10\mu\text{m}$, $y$=$7\mu\text{m}$), P3($x$=$-10\mu\text{m}$, $y$=$10\mu\text{m}$), and P4($x$=$10\mu\text{m}$, $y$=$10\mu\text{m}$), to collect the electric and magnetic fields data. The probe positions are denoted by the filled circles in Figs.~\ref{fig:fig1}(g) and Figs.~\ref{fig:fig1}(j).

\begin{figure}
\includegraphics[scale=0.45]{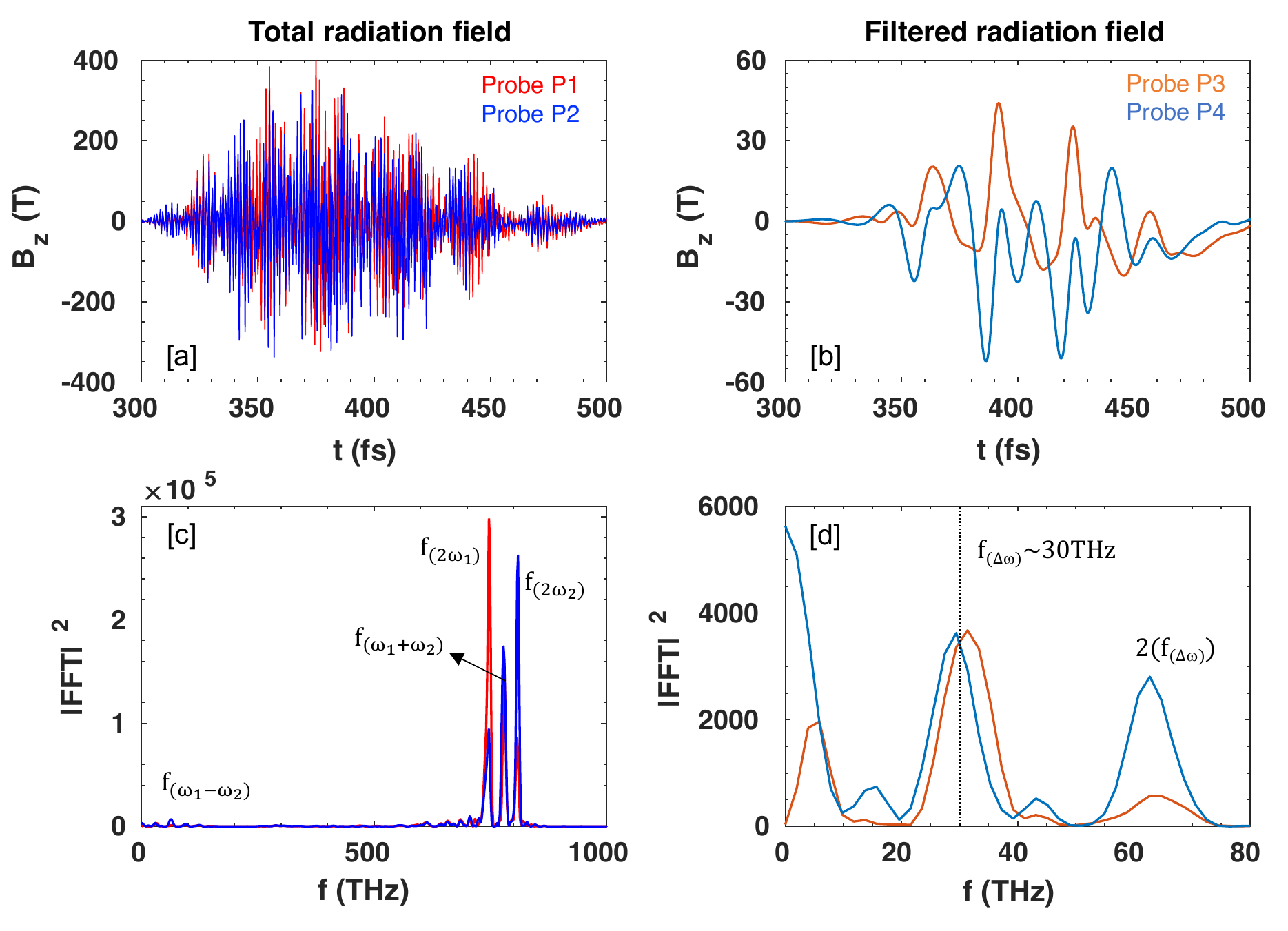}
\caption{\label{fig:fig2} Temporal profiles of 
(a) total emitted radiation,
(b) filtered radiation (beat-frequency) in vacuum, measured by the virtual probes; P1, P2, P3 and P4, located $10\mu\text{m}$ away from the emission spot, and 
(c)-(d) their corresponding Fourier spectra.}
\end{figure}

The temporal evolutions of the total and filtered radiation fields obtained from different probes are shown in Figs.~\ref{fig:fig2}(a)-(b). The Fourier spectra of their temporal dependence have the dominant peaks at $2\omega_{1}$, $2\omega_{2}$ (SH of the laser pulses), $\omega_{1}$$+$$\omega_{2}$, and a small peak at $\omega_{1}$$-$$\omega_{2}$ [Figs.~\ref{fig:fig2}(c)]. Here, we note that the second harmonic is generated by reflection; $2\omega_{1}$ is dominant on the front side of the target,  while  $2\omega_{2}$ on the rear side of the target (indicated by the red and blue color lines, respectively). In this study, we focus only on the beat-frequency radiation, so we have removed the high-frequency components from the obtained spectra; Fig.~\ref{fig:fig2}(d) shows a dominant peak near the beat-frequency ($f_{\Delta\omega}$$\simeq$$30\text{THz}$) (which represents the emission central frequency), and other smaller peaks at its harmonics $2(f_{\Delta\omega})$. They are obtained from the probe (P3) located at the front (left) side of the target. Also, the spectrum has a low-frequency peak near $f$$\simeq$$5\text{THz}$, which we believe comes through the transition radiation by fast electrons generated from laser-target interactions via different mechanisms \cite{Brunel49, Xu50, Bauer51}. A strong DC-like, wideband (0$\sim$10THz) component is observed from the probe (P4) located at the rear (right) side of the target [Fig.~\ref{fig:fig2}(d)], which also comes from transition radiation. Comparable strength of the peak at beat-frequency is also found on the rear side. The peak field strengths are up to $44.2\text{T}$ and $51.4\text{T}$ at $\pm10\mu\text{m}$ from the sheet surface, which is much higher than other THz sources reported so far.

\begin{figure}[h!]
\includegraphics[scale=0.70]{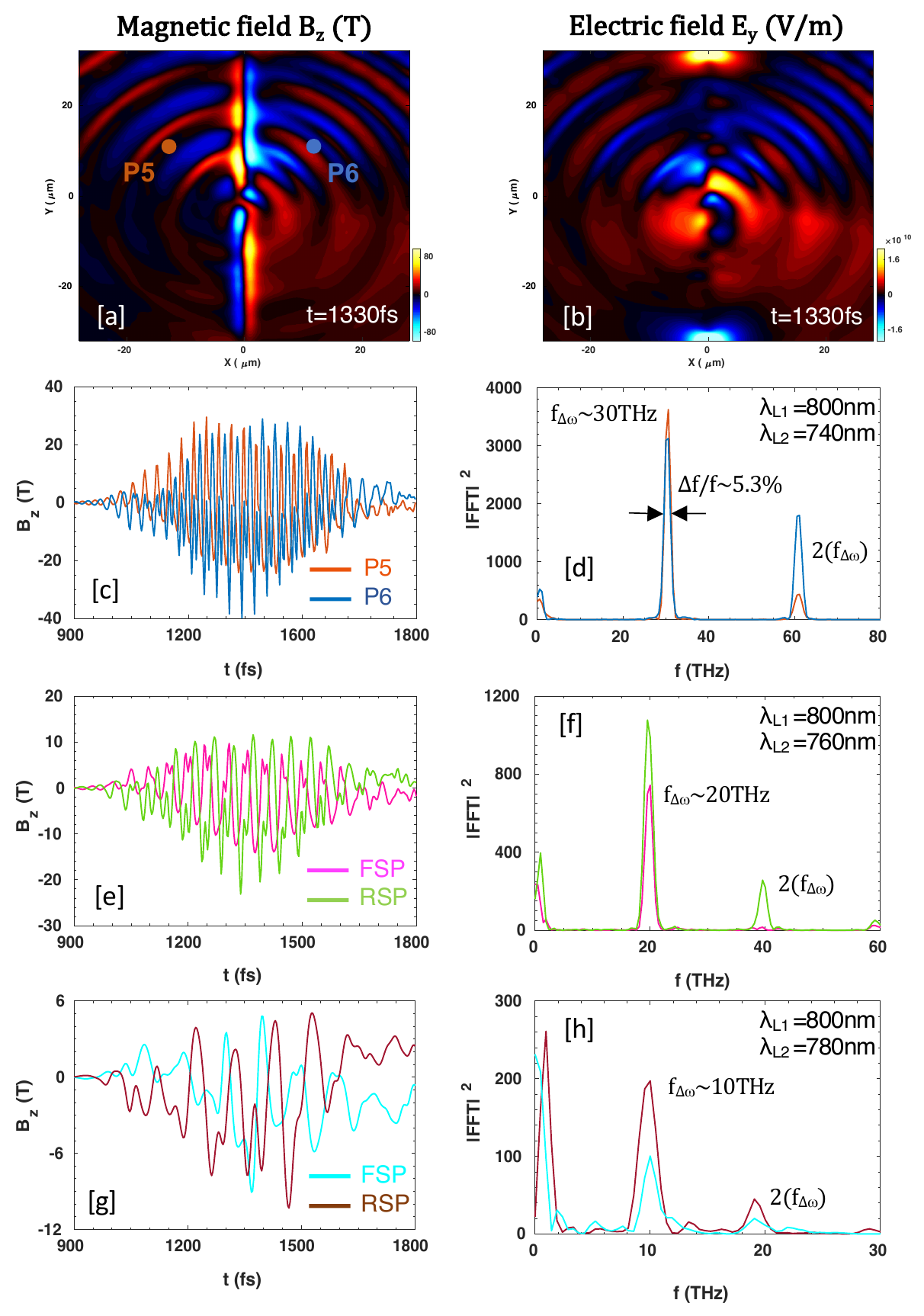}
\caption{\label{fig:fig3}%
Snapshots of spatial distributions of 
(a) magnetic field $B_{z} (\text{T})$ and
(b) electric field $E_{y} (\text{V}/\text{m})$ at $t$=$1330\text{fs}$ for longer pulse duration $\tau_{L}$=$500\text{fs}$. The field data are filtered to extract the beat-frequency THz range. 
(c) Temporal evolutions of filtered magnetic fields which are acquired from probes P5 and P6 (in vacuum), and 
(d) their corresponding power spectra. 
When the wavelength of the second laser pulse is changed, temporal profiles and power spectra; 
(e)-(f) for $\lambda_{L2}$$=$$\text{760nm}$, and 
(g)-(h) for $\lambda_{L2}$$=$$\text{780nm}$.} 
\end{figure}

We now study the cases driven by much longer laser pulses,  with duration $\tau_{L}$$=$$500\text{fs}$ on the same target. In the snapshot of $B_{z}$, strong emission is seen at front and rear surfaces [Fig.~\ref{fig:fig3}(a)]. Interestingly, strong radiation is found at top and bottom edges of the sheet in the snapshot of $E_{y}$ [Fig.~\ref{fig:fig3}(b)], which could be related to antenna-like emission by shielding electron current flowing along the plasma sheet surface \cite{Zhuo52, Dechard53}. Fig.~\ref{fig:fig3}(c) shows the temporal profiles of the magnetic field data collected by the probes; P5($x$=$-10\mu\text{m}$, $y$=$10\mu\text{m}$) and P6($x$=$10\mu\text{m}$, $y$=$10\mu\text{m}$), marked by filled orange and light-blue circles in Fig.~\ref{fig:fig3}(a). It is noteworthy in Fig.~\ref{fig:fig3}(c) that the number of cycles in the radiation pulse has increased in accordance with the increase in pulse duration of the laser. The THz pulse duration is comparable to the laser pulse duration, indicating that THz radiation is generated all through pulse-plasma interaction. Compared to Fig.~\ref{fig:fig2}(d), the bandwidth of the Fourier transform around the beat-frequency ($\simeq$$30\text{THz}$) and its second harmonic ($\simeq$$60\text{THz}$) is considerably narrowed, up to $5.3\%$ [Fig.~\ref{fig:fig3}(d)]. The peak field strength reaches up to $39.8\text{T}$ ($\simeq$$11.9\text{GV/m}$). Comparable strength of peak electric field of $10.8\text{GV/m}$ was reported via conventional difference frequency mixing in AgGaS$_{2}$ \cite{Sell54}. Here, we notice that the second harmonic of the beat-frequency on the right side of the sheet is stronger than on the left side of the sheet. We have not yet identified the reason for that. 

We also performed numerical simulations for the effects of detuning of driving pulses. Figure.~\ref{fig:fig3}(e) represents the THz radiation field measured at the same probe positions as before (front side probe (FSP) and rear side probe (RSP)), when the wavelength of the second laser pulse is increased to $\lambda_{L2}$=$\text{760nm}$ (corresponding to $20\text{THz}$). The radiation field amplitude decreases from the previous case with $\lambda_{L2}$=$\text{740nm}$, as the beat current is proportional to the beat-frequency (actually the linear frequency dependence of radiation amplitude is common for the scheme that uses the plasma electrons as charge carriers).  Its power spectrum has a main peak at the beat-frequency ($\simeq$$20\text{THz}$) [Fig.~\ref{fig:fig3}(f)]. When the wavelength of the second laser pulse is increased even more up to 780nm, the THz radiation again peaks at the beat-frequency, \textit{i.e.} $10\text{THz}$ [Fig.~\ref{fig:fig3}(h)]. In the lower beat-frequency cases also, the THz pulse durations are comparable to the driving laser pulse duration [Fig.~\ref{fig:fig3}(g)], implying that by using longer laser pulses (\textit{e.g.} $\sim$$1\text{ps}$), the number of cycles in the radiation pulse and the beat frequency peak value can be increased. Note that the low-frequency radiations near $f$$\sim$$1\text{THz}$, throughout Fig.~\ref{fig:fig3}(c)-(h), do not dramatically change as the second wavelength of the pulse increases, evidencing that the low-frequency radiation comes dominantly from single-laser-target interactions (\textit{i.e.} CTR) rather than from the beating of the lasers.

\begin{figure}[h!]
\includegraphics[scale=0.42]{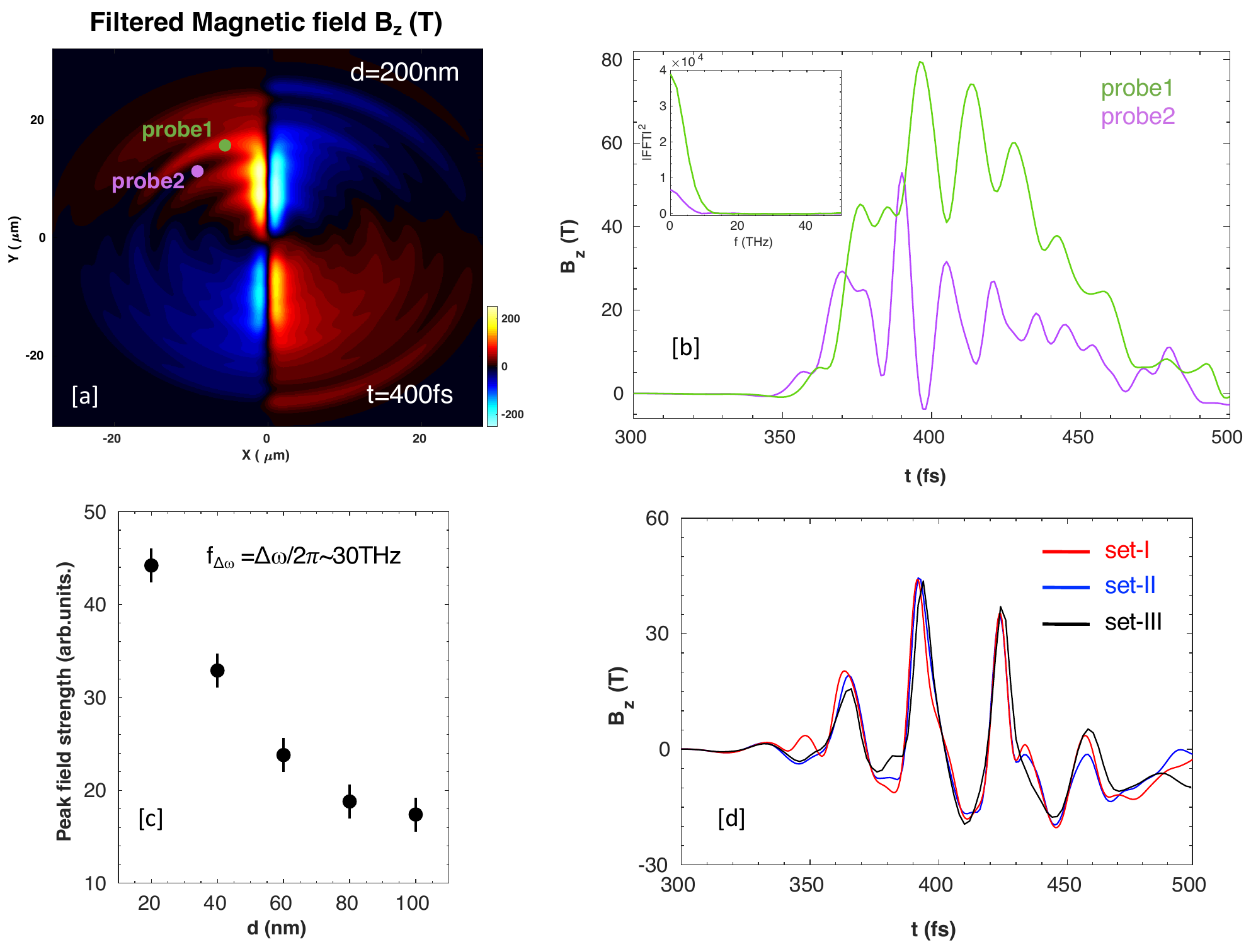}
\caption{\label{fig:fig4} 
(a) Snapshot of spatial distribution of filtered magnetic field $B_{z} (\text{T})$ at $t$=$400\text{fs}$ for target thickness $d$=$200\text{nm}$. 
(b) Temporal profiles of emitted radiation measured by probes; probe1($x$=$-5\mu\text{m}$, $y$=$15\mu\text{m}$) and probe2 ($x$=$-10\mu\text{m}$, $y$=$10\mu\text{m}$).
(c) Peak field strength of beat-frequency radiation as a function of the plasma sheet thickness.
(d) Temporal profiles of emitted radiation measured by front side probe($x$=$-10\mu\text{m}$, $y$=$10\mu\text{m}$) in vacuum for three different sets; set-I ($d$=$20\text{nm}$, $n_{e0}$=$5n_{c}$), set-II ($d$=$10\text{nm}$, $n_{e0}$=$10n_{c}$), and set-III ($d$=$2\text{nm}$, $n_{e0}$=$50n_{c}$).} 
\end{figure}

We also examine the effect of the thickness of plasma sheet. Figure~\ref{fig:fig4}(a) shows the snapshot of filtered radiation field for thickness $d$=$200\text{nm}$, with other parameters same as in Fig.~\ref{fig:fig2}.  Differently from a thin target ($d$=$20\text{nm}$), the radiation fills the whole simulation domain (by contrast, upward radiation is dominant in Fig.~\ref{fig:fig1}(j)). Note that, it has different polarities at the front and back surfaces of the sheet. Radiation emission to the surface-normal direction is very small, i.e. conical emission in 3D space, typical of CTR. The temporal profiles of emitted radiation in vacuum and their spectra are illustrated in Fig.~\ref{fig:fig4}(b). From a thick target ($d$=$200\text{nm}$), only the half-cycle, low-frequency radiation is generated; no beat-frequency radiation is observed. This is because the target is thicker than the skin depth, so the laser pulses cannot beat with each other. The target interacts with single pulses from either side and the radiation can be produced via CTR mechanism only. Figure.~\ref{fig:fig4}(c) shows the peak field strength of beat-frequency THz radiation decreases with increasing the thickness of the plasma sheet.
We also studied the effect of target density. For all the sets; set-I, set-II, and set-III, we preserved the surface density to 100 (nm $\times n_{c}$), and other parameters are the same as in Fig.~\ref{fig:fig2}. The temporal profiles of the filtered radiation field in vacuum are illustrated in Fig.~\ref{fig:fig4}(d). Here, we find that the beat-frequency emitted radiation changes only slightly for different thicknesses but with the same surface density. This result, along with the insensitivity to the density profile of the target, can be a great advantage in designing future experiments; one may use a free-diffused target starting from the ablation of high-density, but thin targets. In this case, the emission of THz may be insensitive to the delay between the ablating pulse and driver pulses. Other is the use of a graphene sheet. Furthermore, our 2D simulation results also suggest that the system (neutral target of thickness=$2$nm with density=50$n_{c}$) can be produced the beat-frequency THz radiation with high peak field strength. 

\begin{figure}[h!]
\includegraphics[scale=0.40]{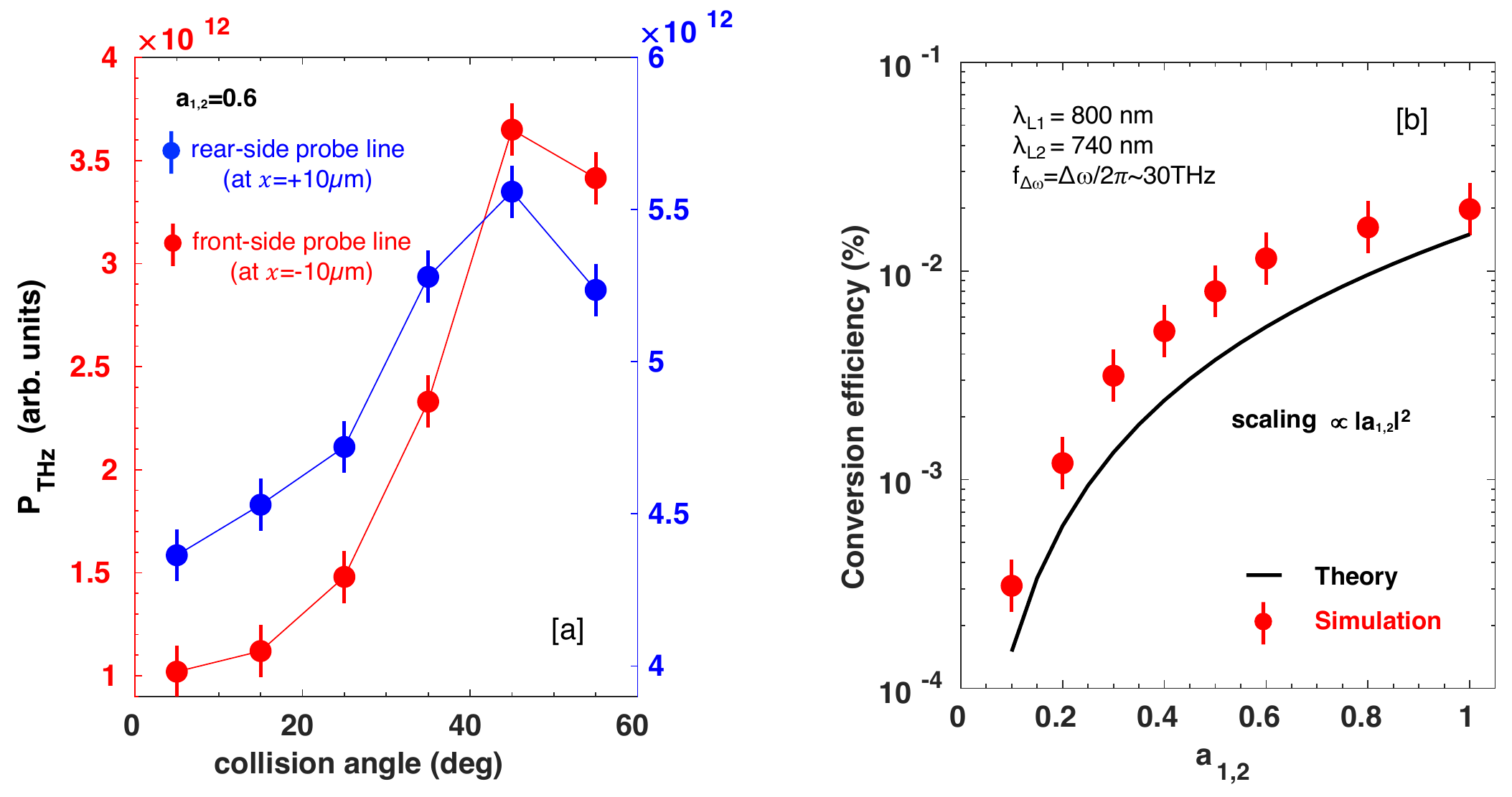}
\caption{\label{fig:fig5} 
(a) Power of THz emission into vacuum vs. colliding angles of the laser pulses. The power is obtained by integrating over $y$ (from $-32\mu \text{m}$ through $+32\mu \text{m}$) of the THz emission passing through the left and right side probe lines at $x=\pm 10\mu \text{m}$.
(b) The efficiency of power conversion from lasers-to-THz emissions as a function of normalized vector potential of lasers $a_{1,2}$. The solid red circles are from simulation results (measured at $10\mu \text{m}$ from the emission spot), and the solid black line is from theoretical results.}
\end{figure}

Now we see the behaviour of THz radiation yield for different colliding angles of the laser pulses. The average Poynting flux ($<$$\text{S}_{rad}$$>$=$(c/2\mu_{0})|B_{0}|^{2}$, $B_{0}$ is the peak value of magnetic field $B_{z}$ and $\mu_{0}$ is the free space permeability) passing through the probing planes (actually probing lines in 2D domain) at $x$=$\pm10\mu\text{m}$ (both left and right sides of the sheet) is integrated over $y$ from $y_{1}$=$-32\mu\text{m}$, through $y_{2}$=$+32\mu\text{m}$, to obtain the power of THz radiation. Fig.~\ref{fig:fig5}(a) shows that the peak power of beat-frequency THz radiation increases as the collision angle increases. This is because the transverse current (or emission source area) is increased for the larger collision angles, resulting in an enhancement in the field strength of radiation pulses.  However, for collision angle $\theta_{\text{col}}$$>$$50^\circ$, it eventually saturates as the emission is confined near the target surfaces. Note that the direct-current components also decrease as the collision angle increases.

In our approach, THz radiation arises from the strong plasma current generated when two laser pulses collide with each other on a nano-dimensional overdense plasma sheet target. During the interaction, both longitudinal and transverse currents are formed, carried by the plasma electrons. These nonlinear current sources can be treated like a wire antenna in this limit ($d$$<<$$\lambda_{\text{THz}}$), as far as the radiation field is concerned \cite{Kumar55}. The calculated total radiated THz power can be written by;
\begin{eqnarray}
{\text{P}_{\text{THz}}}=&&\frac{\pi cm^{2}\omega_{p}^{4}\sin^{2}\theta_{\text{col}}}{2\mu_{0}e^{2}}\left(\frac{a_{1}a_{2}^\ast}{2}\right)^{2}\nonumber\\
&&\times
\int_{0}^{\pi}{\frac{\sin^{2}\left[ {\Delta\omega}L(1-\cos{\theta})/2c \right] \sin^{3}{\theta}I_{0}^{2}}{(1-\cos{\theta})^{2}}d\theta},
\label{eq:six}   
\end{eqnarray}
where $I_{0}=\int_{0}^{d}{J_{0}(\Delta\omega \rho_i \sin{\theta}/c)\rho_i d\rho_i}$, $J_{0}(x)$ is the Bessel function of first kind zero order, $L$ and $d$ are the length and radius of the wire, $\omega_{p}^{2}$$=$$n_{0}e^{2}/m\varepsilon_{0}$, is the plasma frequency, $a_{1}$$=$$eA_{1}/m\omega_{1}c$ and $a_{2}^\ast$$=$$eA_{2}^\ast/m\omega_{2}c$ are the normalized amplitude of first and second laser pulses. In this calculation, we have considered the radiation driven by the beat-frequency ponderomotive force only. The above equation predicts that the emission power increases with $\sin^{2}\theta_{\text{col}}$ for $\theta_{\text{col}} $$<$${90}^\circ$. There is no THz emission for normal incidence $\theta_{\text{col}}$$=$$0^\circ$ because there is no transverse current component. 
Fig.~\ref{fig:fig5}(b) shows the lasers-to-THz power conversion efficiency as a function of the normalized vector potential of lasers $(a_{1,2})$, and a comparison of the theoretically obtained efficiency (from Eq.~(\ref{eq:six})) with simulation results. The efficiency increases with $a_{1,2}$ and its value reaches around $0.02\%$ when the laser intensities $>$$10^{18}\text{W}/\text{cm}^{2}$.  However, the theoretically obtained efficiency is small compared to simulation results, because the longitudinal current is not included in THz power calculations (only considered the transverse current), and its scaling is proportional to the square of the normalized vector potential of lasers.

In summary, we proposed a novel scheme to obtain narrowband, readily tunable, compact, and experiment-friendly THz sources based on the oblique collision of two-detuned laser pulses on a nano-dimensional overdense plasma sheet target. The process of THz emission and resulting THz characteristics were investigated by using two-dimensional particle-in-cell simulations. In contrast to single-pulse-driven targets, where long half-cycle THz pulses are generated with a broadband spectrum via CTR mechanism, by irradiating the rear surface of the target with another laser pulse, we produced multicycle, narrowband THz radiation at the beating frequency, dominating over the CTR. The THz radiation was generated due to the ponderomotive force-driven plasma current in the colliding region. The number of cycles of THz oscillation can be increased arbitrarily by increasing the driving pulse duration and the system can provide a very narrowband spectrum. We obtained extremely intense THz radiation of 40$\text{TW}/\text{cm}^{2}$, with peak fields up to  $\sim$11.9GV/m and 39.8T at 10$\mu$m from the emission spot, with central frequency $f$$\sim$$30\text{THz}$ and spectral bandwidth $\Delta f/f$$\simeq$$5.3\%$ for $\tau_{L}$$=$$500\text{fs}$. For the laser intensities $>$$10^{18}\text{W}/\text{cm}^{2}$, the emitted THz pulse energy was about 0.1mJ. Such an extremely intense narrowband THz source will be suitable for various ambitious applications such as compact electron accelerators \cite{Nanni44} and pump-probe experiments \cite{Forst45, Dienst46, Bakke47}. We believe our results will help to interpret future experiments in this field.

This research work was supported by the National Research Foundation (NRF) of Korea (Grant Nos. NRF-2.220786.01, NRF-2020R1A2C1102236, NRF-2016R1A5A1013277, and NRF-2022R1A2C3013359).

\begin{thebibliography}{9}
\bibitem{Tonouchi1}M. Tonouchi, {Nat. Photonics} \textbf{1}, 97 (2007).
\bibitem{Zhang2}X. C. Zhang, A. Shkurinov, and Y. Zhang, {Nat. Photonics} \textbf{11}, 16 (2017).
\bibitem{Pickwell3}E. Pickwell and V. Wallace, {J. Phys. D Appl. Phys.} \textbf{39}, R301 (2006).
\bibitem{Mittleman4}D. M. Mittleman, {Opt. Express} \textbf{26}, 9417 (2018).
\bibitem{Jepsen5}P. U. Jepsen, D. G. Cooke, and M. Koch, {Laser Photon. Rev.} \textbf{5}, 124 (2011).
\bibitem{Baxter6}J. B. Baxter and G. W. Guglietta, {Laser Photon. Rev.} \textbf{5}, 124 (2011).
\bibitem{Liu7}J. Liu, J. Dai, S. L. Chin, and X. C. Zhang, {Nat. Photonics} \textbf{4}, 627 (2010).
\bibitem{Koenig8}S. Koenig, D. Lopez-Diaz, J. Antes, F. Boes, R. Henneberger, A. Leuther, A. Tessmann, R. Schmogrow, D. Hillerkuss, R. Palmer, T. Zwick, C. Koos, W. Freude, O. Ambacher, J. Leuthold, and I. Kallfass, {Nat. Photonics} \textbf{7}, 977 (2013).
\bibitem{Kemp9}M. Kemp, {IEEE Trans. THz Sci. Technol.} \textbf{1}, 282 (2011).
\bibitem{Auston10}D. H. Auston, K. Cheung, J. Valdmanis, and D. Kleinman, {Phys. Rev. Lett.} \textbf{53}, 1555 (1984).
\bibitem{Stepanov11}G. A. Stepanov, L. Bonacina, S. V. Chekalin, and J. P. Wolf, {Opt. Lett.} \textbf{33}, 2497 (2008).
\bibitem{Fulop12}J. A. Fulop, Z. Ollmann, C. Lombosi, C. Skrobol, S. Klingebiel, L. Palfalvi, F. Krausz, S. Karsch,  and J. Hebling, {Opt. Express} \textbf{22}, 20155 (2014).
\bibitem{Huang13}W. Huang, S. Huang, E. Granados, K. Ravi, K. Hong, L. Zapata, and F. Kartner, {J. Mod. Opt.} \textbf{62}, 1486 (2015).
\bibitem{Vicario14}C. Vicario, A. Ovchinnikov, S. Ashitkov, M. Agranat, V. Fortov, and C. Hauri, {Opt. Lett.} \textbf{39}, 6632 (2014).
\bibitem{Vicario15}C. Vicario, B. Monoszlai, and C. P. Hauri, {Phys. Rev. Lett.} \textbf{112}, 213901 (2014).
\bibitem{Hamster16}H. Hamster, A. Sullivan, S. Gordon, W. White, and R. W. Falcone, {Phys. Rev. Lett.} \textbf{71}, 2497 (1993).
\bibitem{Sheng17}Z. M. Sheng, K. Mima, J. Zhang, and H. Sanuki, {Phys. Rev. Lett.} \textbf{94}, 95003 (2005).
\bibitem{Kumar18}M. Kumar, K. Lee, S. H. Park, Y. U. Jeong, and N. Vinokurov,  {Phys. Plasmas} \textbf{24}, 033104 (2017).
\bibitem{Kwon19}K. B. Kwon, T. Kang, H. S. Song, Y. K. Kim, B. Ersfeld, D. A. Jaroszynski, and M. S. Hur, {Sci. Rep.} \textbf{8}, 145 (2018).
\bibitem{Kumar20}M. Kumar, T. Kang, S. Kylychbekov, H. S. Song, and M. S. Hur, {Phys. Plasmas} \textbf{28}, 033101 (2021).
\bibitem{Amico21}C. D. Amico, A. Houard, M. Franco, B. Prade, A. Mysyrowicz, A. Couairon, and V. T. Tikhonchuk, {Phys. Rev. Lett.} \textbf{98}, 235002 (2007).
\bibitem{Kim22}K. Y. Kim, J. H. Glownia, A. J. Taylor, and G. Rodriguez, {Opt. Express} \textbf{15}, 4577 (2007).
\bibitem{Dai23}J. Dai, X. Xie, and X. C. Zhang, {Phys. Rev. Lett.} \textbf{97}, 103903 (2006).
\bibitem{Liu24}K. Liu, A. D. Koulouklidis, D. G. Papazoglou, S. Tzortzakis, and X. C. Zhang, {Optica} \textbf{3}, 605 (2016).
\bibitem{Kuk25}D. Kuk, Y. J. Yoo, E. W. Rosenthal, N. Jhajj, H. M. Milchberg, and K. Y. Kim, {Appl. Phys. Lett.} \textbf{108}, 121106 (2016).
\bibitem{Leemans26}W. P. Leemans, C. Geddes, J. Faure, C. Toth, J. V. Tilborg, C. B. Schroeder, E. Esarey, G. Fubiani, D. Auer- bach, B. Marcelis, M. A. Carnahan, R. A. Kaindl, J. Byrd, and M. C. Martin, {Phys. Rev. Lett.} \textbf{91}, 074802 (2003).
\bibitem{Schroeder27}C. B. Schroeder, E. Esarey, J. V. Tilborg, and W. P. Leemans, {Phys. Rev. E} \textbf{69}, 016501 (2004).
\bibitem{Liao28}G. Q. Liao, Y. T. Li, Y. H. Zhang, H. Liu, X. L. Ge, S. Yang, W. Q. Wei, X. H. Yuan, Y. Deng, B. J. Zhu, Z. Zhang, W. M. Wang, Z. M. Sheng, L. M. Chen, X. Lu, J. L. Ma, X. Wang, and J. Zhang, {Phys. Rev. Lett.} \textbf{116}, 205003 (2016).
\bibitem{Liao29}G. Q. Liao, Y. T. Li, C. Li, H. Liu, Y. H. Zhang, W. M. Jiang, X. H. Yuan, J. Nilsen, T. Ozaki, W. M. Wang, Z. M. Sheng, D. Neely, P. McKenna, J. Zhang, {Plasma Phys. Control. Fusion} \textbf{59}, 014039 (2017).
\bibitem{Dechard30}J. Dechard, X. Davoine, and L. Berge, {Phys. Rev. Lett.} \textbf{123}, 264801 (2019).
\bibitem{Ding31}W. J. Ding, Z. M. Sheng, and W. S. Koh,{App. Phys. Lett.} \textbf{103}, 204107 (2013).
\bibitem{Gopal32}A. Gopal, S. Herzer, A. Schmidt, P. Singh, A. Reinhard, W. Ziegler, D. Brommel, A. Karmakar, P. Gibbon, U. Dillner, T. May, H. G. Meyer, and G. G. Paulus, {Phys. Rev. Lett.} \textbf{111}, 074802 (2013).
\bibitem{Gopal33}A. Gopal, T. May, S. Herzer, A. Reinhard, S. Minardi, M. Schubert, U. Dillner, B. Pradarutti, J. Polz, T. Gaumnitz, M. C. Kaluza, O. Jackel, S. Riehemann, W. Ziegler, H. P. Gemuend, H. G. Meyer and G. G. Paulus, {New J. Phys.} \textbf{14}, 083012 (2012).
\bibitem{Li34}Y. T. Li, C. Li, M. L. Zhou, W. M. Wang, F. Du, W. J. Ding, X. X. Lin, F. Liu, Z. M. Sheng, X. Y. Peng, L. M. Chen, J. L. Ma, X. Lu, Z. H. Wang, Z. Y. Wei, and J. Zhang, {Appl. Phys. Lett.} \textbf{100}, 254101 (2012).
\bibitem{Li35}C.Li,M.L.Zhou,W.J.Ding,F.Du,F.Liu,Y.T.Li, W. M. Wang, Z. M. Sheng, J. L. Ma, L. M. Chen, X. Lu, Q. L. Dong, Z. H. Wang, Z. Lou, S. C. Shi, Z. Y. Wei, and J. Zhang, {Phys. Rev. E} \textbf{84}, 036405 (2011).
\bibitem{Jin36}Z. Jin, H. B. Zhuo, T. Nakazawa, J. H. Shin, S. Waka- matsu, N. Yugami, T. Hosokai, D. B. Zou, M. Y. Yu, Z. M. Sheng, and R. Kodama, {Phys. Rev. E} \textbf{94}, 033206 (2016).
\bibitem{Herzer37}S. Herzer, A. Woldegeorgis, J. Polz, A. Reinhard, M. Almassarani, B. Beleites, F. Ronneberger, R. Grosse, G. G. Paulus, U. Hubner, T. May and A. Gopal, {New J. Phys.} \textbf{20}, 063019 (2018).
\bibitem{Liao38}G. Q. Liao, H. Liu, G. G. Scott, Y. H. Zhang, B. J. Zhu, Z. Zhang, Y. T. Li, C. Armstrong, E. Zemaityte, P. Brad- ford, D. R. Rusby, D. Neely, P. G. Huggard, P. McKenna, C. M. Brenner, N. C. Woolsey, W. M. Wang, Z. M. Sheng, and J. Zhang, {Phys. Rev. X} \textbf{10}, 031062 (2020).
\bibitem{Liao39}G. Q. Liao, Y. T. Li, H. Liu, G. G. Scott, D. Neely, Y. H. Zhang, B. J. Zhu, Z. Zhang, C. Armstrong, E. Zemaityte, P. Bradford, P. G. Huggardi, D. R. R. nd P. McKenna, C. M. Brenner, N. C. Woolsey, W. M. Wang, Z. M. Sheng, and J. Zhang, {Proc. Natl. Acad. Sci.} \textbf{116}, 3994 (2019).
\bibitem{Mondal40}S. Mondal, Q. W. W. J. Ding, H. A. Hafez, M. A. Fareed, A. Laramee, X. Ropagnol, G. Zhang, S. Sun, Z. M. Sheng, J. Zhang, and T. Ozaki, {Sci. Rep.} \textbf{7}, 40058 (2017).
\bibitem{Tokita41}S. Tokita, S. Sakabe, T. Nagashima, M. Hashida, and S. Inoue, {Sci. Rep.} \textbf{5}, 8268 (2015).
\bibitem{Zhang42}J. Zhang, X. Ban, F. Wan, and C. Lv, {Appl. Sci.} \textbf{12}, 4464 (2022).
\bibitem{Spies43}J. A. Spies, J. Neu, U. T. Tayvah, M. D. Capobianco, B. Pattengale, S. Ostresh, and C. A. Schmuttenmaer,{J. Phys. Chem. C} \textbf{124}, 22335 (2020).
\bibitem{Nanni44}E. A. Nanni, W. R. Huang, K. H. Hong, K. Ravi, A. Fallahi, G. Moriena, R. J. D. Miller, and F. X. Kartner, {Nat. Commun.} \textbf{6}, 8486 (2015).
\bibitem{Forst45}M. Forst, C. Manzoni, S. Kaiser,  Y. Tomioka, Y. Tokura, R. Merlin, and A. Cavalleri, {Nat. Phys.} \textbf{7}, 854 (2011).
\bibitem{Dienst46}A. Dienst, M. C. Hoffmann, D. Fausti, J. C. Petersen, S. Pyon, T. Takayama, H. Takagi, and A. Cavalleri, {Nat. Photonics} \textbf{5}, 485 (2011).
\bibitem{Bakke47}D. J. Bakker, A. Peters, V. Yatsyna, V. Zhaunerchyk, and A. M. Rijs, {J. Phys. Chem. Lett.} \textbf{7}, 1238 (2016).
\bibitem{Arber48}T. D. Arber, K. Bennett, C. S. Brady, A. L. Douglas, M. G. Ramsay, N. Sircombe, P. Gillies, R. G. Evans, H. Schmitz, A. R. Bell, and C. P. Ridgers, {Plasma Phys. Control. Fusion} \textbf{57}, 113001 (2015).
\bibitem{Brunel49}F. Brunel, {Phys. Rev. Lett.} \textbf{59}, 52 (1987).
\bibitem{Xu50}H. Xu, Z. M. Sheng, J. Zhang, and M. Y. Yu, {Phys. Plasmas} \textbf{13}, 123301 (2006).
\bibitem{Bauer51}D. Bauer and P. Mulser, {Phys. Plasmas} \textbf{14}, 023301 (2007).
\bibitem{Zhuo52}H. B. Zhuo, S. J. Zhang, X. H. Li, H. Y. Zhou, X. Z. Li, D. B. Zou, M. Y. Yu, H. C. Wu, Z. M. Sheng, and C. T. Zhou, {Phys. Rev. E}. \textbf{95}, 013201 (2017).
\bibitem{Dechard53}J. Dechard, X. Davoine, L. Gremillet, and L. Berge, {Phys. Plasmas} \textbf{27}, 093105 (2020).
\bibitem{Sell54}A. Sell, A. Leitenstorfer, and R. Huber, {Opt. Lett.} \textbf{33}, 2767 (2008).
\bibitem{Kumar55}M. Kumar and V. K.Tripathi, {Phys. Plasmas} \textbf{18}, 053105 (2011).
\end{thebibliography}

\end{document}